\documentstyle[epsfig,aps,floats]{revtex}
\begin{document}
\draft
\preprint{\parbox{6cm}{\flushright \today \\UAB--FT--464
\\FERMILAB--Pub--99/066--A\\[0.6cm]}}
\title{Photon Spectrum Produced by \\ the Late Decay of a
Cosmic Neutrino Background}
\author{Eduard Mass\'o}
\address{Grup de F\'{\i}sica Te\`orica and Institut de F\'{\i}sica
d'Altes Energies \\ Universitat Aut\`onoma de Barcelona \\
08193 Bellaterra, Catalonia, Spain}
\author{Ramon Toldr\`a}
\address{NASA/Fermilab Astrophysics Group \\ 
         Fermi National Accelerator Laboratory \\
         Box 500, Batavia, IL 60510-0500, USA}
\maketitle
\begin{abstract}
We obtain the photon spectrum induced by a cosmic background of 
unstable neutrinos. We study the spectrum in a variety of cosmological
scenarios and also we allow for the neutrinos having a momentum
distribution (only a critical matter dominated universe and neutrinos
at rest have been considered until now). Our results can be helpful
when extracting bounds on neutrino electric and magnetic moments
from cosmic photon background observations.
\end{abstract}
\vspace{1cm}

\section{Introduction}

Recent indications for neutrino masses 
\cite{SuperK}
have strengthened the interest in physical effects linked to
massive neutrinos. In this paper we shall be concerned with one of these
effects, namely, with the fact that massive neutrinos can have non--zero
electromagnetic form factors and can be unstable due to radiative decay
processes. 

Specifically, we shall be interested in the contribution of decays
$\nu_i\rightarrow\nu_j\gamma$, when coming from a cosmological $\nu_i$
density, to the cosmic photon background. A previous study of such
phenomena was performed by Ressell and Turner \cite{ReTu} (see also
\cite{KoTu}). These authors considered the photon background at
different photon energies and constrained the radiative decay of a
cosmic density of massive neutrinos. Recently, Biller {\it et al.}
\cite{Biller} have improved substantially the constraints in the
infrared region by making use of the effect of $\gamma-\gamma$
interactions giving electron--positron pairs, where one $\gamma$ is a
background infrared photon, and the other is a TeV $\gamma$ coming
from an active galaxy. We may expect further improvements in the
future, along the lines used in \cite{Biller} or perhaps with some
new ideas and/or observations.

Constraints on the neutrino lifetime lead to limits on the magnetic
and electric transition moments, $\mu_{ij}$ and $\epsilon_{ij}$, of
neutrinos. The neutrino lifetime $\tau$ is related to these parameters, 
\begin{equation}\label{B}
{B\over\tau}=
{{|\mu_{ij}|^2 + |\epsilon_{ij}|^2}\over {8 \pi}}\,
\left(  {\Delta m^2} \over m_i \right)^3
\end{equation}
where $\Delta m^2=m_i^2 - m_j^2$,
and $B$ is the branching ratio $BR(\nu_i\rightarrow\nu_j\gamma)$.

All the studies performed until now make the simplifying assumptions
that the decays are produced 1) in a critical matter dominated
universe and 2) by neutrinos at rest. Although this might be enough
for the results obtained until now, in the future it may be
interesting to study the contribution to the photon background with
more generality. In a more general framework we can be able to see how
the limits on the neutrino lifetime depend on different assumptions,
and we can extract more precise and reliable bounds. Also, in the
eventual case of a positive signal that may come from decays of relic
neutrinos, the general study of the consequences of the decay that we
perform in the present paper would become quite necessary. Apart from
the above mentioned references, other papers treating unstable neutrinos
in a cosmological context are listed in \cite{refpapers}.

Non--vanishing values of $\mu_{ij}$ and/or $\epsilon_{ij}$ lead to
other potential effects besides the ones considered in this paper.
For example, plasmon decay into neutrinos in stellar media is
constrained by arguments of stellar energy loss. This leads, in
general, to limits that are quite stringent~\cite{Raffelt}. We would
like to point out that when neutrinos are close to a degeneration in
mass, the limits obtained using Eq.~(\ref{B}) may be relevant since in
this case $\Delta m^2/m_i \ll 1$. In any case, the spirit of the
present paper is not phenomenological. Rather, as we said, we would
like to generalize previous studies on the subject.

The organization of the paper is as follows. In Sec.~\ref{rest} we
examine the contribution to the photon background of the decay
$\nu_i\rightarrow\nu_j\gamma$ in different cosmological scenarios,
keeping the assumption that the neutrinos $\nu_i$ decay at rest. In
Sec.~\ref{moving} we drop this assumption and study the decays of
neutrinos with a momentum distribution. We devote Sec.~\ref{concl} to
the conclusions. Some technical details are developed in two
appendices.

\section{Decaying neutrinos at rest} \label{rest}

In this section we shall calculate the photon spectrum produced by the decay
of a cosmic background of neutrinos at rest. We shall consider the two--body
decay of a neutrino with mass $m_i$ into a photon and a neutrino with mass
$m_j$, with $m_i > m_j$: $\nu_i\rightarrow \nu_j + \gamma$.
The subscripts $i,j=1,2,3$ stand for any neutrino mass
eigenstate, which is a linear combination of the three weak eigenstates
$\nu_e,\nu_\mu,\nu_\tau$.
When a neutrino decays at rest the photon energy is given by
\begin{equation} \label{energy}
\epsilon_0 = \frac{\Delta m^2}{2m_i}, 
\end{equation}
with $\Delta m^2 \equiv m_i^2-m_j^2$. The cosmic expansion redshifts the
photon energy. A photon that at present has energy
$E$ was produced at a redshift $z_0$ given by
\begin{equation} \label{zzero}
1+z_0 = \frac{\epsilon_0}{E}.
\end{equation}
Let $F_E$ be the present energy flux of photons with energy $E$ produced 
by neutrino decay. The flux per unit energy and solid angle is given by
\begin{equation}
\frac{d^2F_E}{dE\, d\Omega} = E\, \frac{d^2F_n}{dE\, d\Omega}, 
\end{equation} 
where $F_n$ is the particle flux at present. It is related to the particle
flux at emission time by the relation
\begin{equation}
\frac{d^2F_n}{d\Omega} = \frac{1}{(1+z_0)^3} 
	\frac{d^2F_n}{d\Omega} (t(z_0)) = 
	\frac{1}{4\pi} \frac{1}{(1+z_0)^3} \delta n_\gamma (t(z_0)),
\end{equation}
where we have included the factor of dilution $(1+z_0)^{-3}$ produced by
the expansion of the universe.
The photon density emitted at $z_0$ is given by the usual decay law
\begin{equation}
\delta n_\gamma (t(z_0)) = B \frac{\delta t}{\tau} n_{\nu_i} (t(z_0)),
\end{equation}
where $\tau$ is the neutrino lifetime and $B$ is the branching
ratio for the radiative decay. For a fixed emission
time $\delta t = H^{-1}(z_0) dE/E$, where $H(z)$ is the Hubble expansion
rate at time $t(z)$. Writing everything together we obtain
\begin{equation} \label{prerest}
\frac{d^2F_E}{dE\, d\Omega} = \frac{1}{4\pi} 
	\frac{n_{\nu_i}(t(z_0))}{(1+z_0)^3}
	\frac{B}{\tau H(z_0)}.
\end{equation}
Choosing a time $t_p\ll \tau$, otherwise arbitrary, and calling the
expansion age of the universe~$t_0$, we can write
\begin{equation} \label{ntilde}
\frac{n_{\nu_i}(t(z_0))}{(1+z_0)^3} = \frac{n_{\nu_i}(t_p)}{(1+z(t_p))^3}
	\exp -\frac{t(z_0)-t_p}{\tau} \equiv
	\tilde{n}_{\nu_i}(t_0) \exp -\frac{t(z_0)-t_p}{\tau},
\end{equation}
where $\tilde{n}_{\nu_i}(t_0)$ would be the present number density of
neutrinos if they did not decay. Taking $t_p=0$ and substituting 
Eq.~(\ref{ntilde}) into Eq.~(\ref{prerest}), we finally obtain
\begin{equation} \label{restspectr}
\frac{d^2F_E}{dE\, d\Omega} = \frac{B}{4\pi} \frac{\tilde{n}_{\nu_i}(t_0)}
	{\tau H(z_0)} \exp -\frac{t(z_0)}{\tau},
\end{equation}
for any $E < \epsilon_0$.
For $E > \epsilon_0$ the photon flux vanishes because photons cannot
be produced with energy larger than $\epsilon_0$ by neutrinos decaying
at rest. From now on we shall set $B=1$. The expansion time
$t(z)$ is given in terms of $H(z)$ by the following integral
\begin{equation} \label{time}
t(z) = \int_0^{t(z)}\! dt =
	\int_z^\infty \! \frac{dz'}{1+z'}\frac{1}{H(z')}.
\end{equation}

Equation~(\ref{restspectr}) is our final expression for the photon
energy flux per unit energy and solid angle produced by the decay of a
cosmic background of neutrinos with negligible velocities. It holds
for any isotropic and homogeneous universe and for any equation of
state for the cosmic fluid.

\begin{figure}[bht]
\begin{center}
\epsfig{file=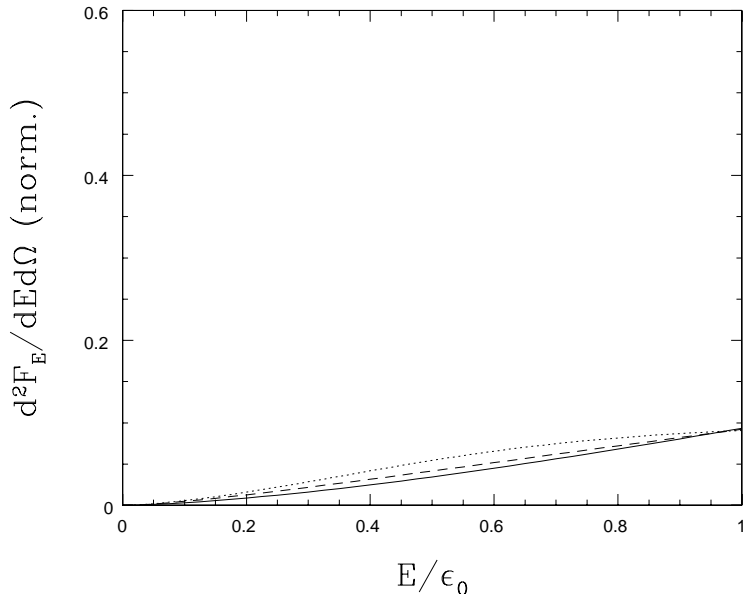,width=10cm,height=10cm}
\caption[]{We plot the normalized spectrum
$(d^2F_E/dE\,d\Omega)/(\tilde{n}_{\nu_i}(t_0)/4\pi)$ versus $E/\epsilon_0$
when $\tau H_0 = 10$. The solid line represents the cosmological model
$\Omega_0=1$ and $\Omega_\Lambda=0$. The dashed line is the model
$\Omega_0=0.3$ and $\Omega_\Lambda=0$. The dotted line is
$\Omega_0=0.3$ and $\Omega_\Lambda=0.7$.}
\label{H0tau100}
\end{center}
\end{figure}

\begin{figure}[bht]
\begin{center}
\epsfig{file=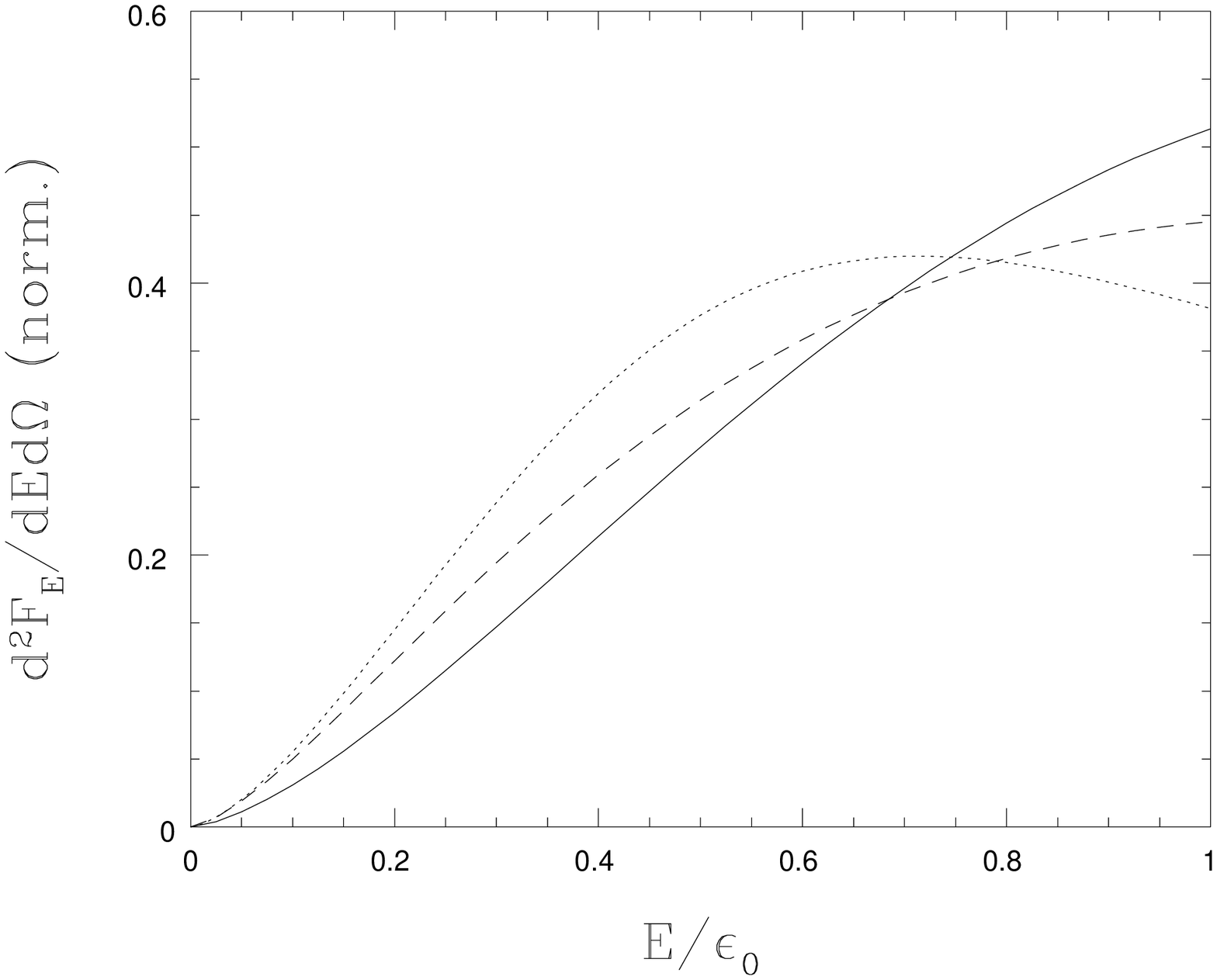,width=10cm,height=10cm}
\caption[]{Same as Fig.~\ref{H0tau100} but with $\tau H_0 = 1$}
\label{H0tau10}
\end{center}
\end{figure}

\begin{figure}[bht]
\begin{center}
\epsfig{file=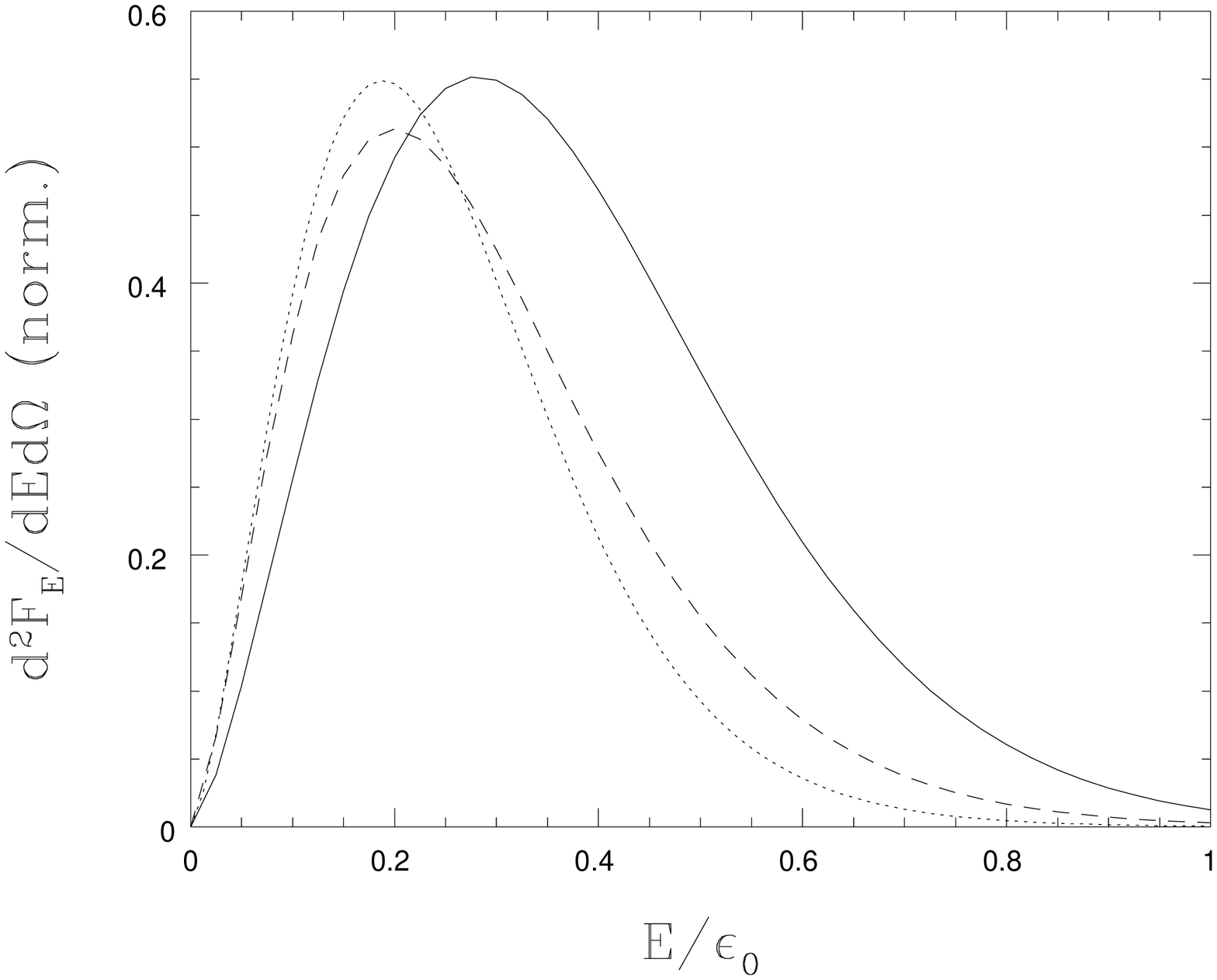,width=10cm,height=10cm}
\caption[]{Same as Fig.~\ref{H0tau100} but with $\tau H_0 = 0.1$}
\label{H0tau01}
\end{center}
\end{figure}

Once the expansion rate $H(z)$ is given, one can calculate $t(z)$ with
Eq.~(\ref{time}) and then calculate $H(z_0)$ and $t(z_0)$ using
Eq.~(\ref{zzero}). Plugging the result into Eq.~(\ref{restspectr}) one
obtains the photon spectrum as a function of the photon energy $E$.
As examples we shall consider three different cosmological scenarios: a
power--law expansion and two observationally more motivated scenarios,
an open universe filled only with matter and a flat universe with
matter and a nonzero cosmological constant $\Lambda$.

First, let us take a power--law expansion. The Hubble expansion rate
and the expansion time are given, respectively, by
\begin{eqnarray}
H(z) &=& H_0(1+z)^n, \label{Hpower}\\
H_0 t(z) &=& \frac{1}{n} \frac{1}{(1+z)^n} \label{tpowerlaw}.
\end{eqnarray} 
with $n\geq 0$ and $H_0$ the Hubble constant \cite{note}.
Particular examples include
a critical $\Lambda$ dominated universe ($n=0$), an empty 
open universe ($n=1$),
a flat matter dominated universe ($n=3/2$), and a flat radiation 
dominated universe ($n=2$). For a critical matter dominated universe
we obtain
\begin{equation}
\frac{d^2F_E}{dE\, d\Omega} = \frac{1}{4\pi} \frac{\tilde{n}_{\nu_i}(t_0)}
	{\tau H_0} \left(\frac{E}{\epsilon_0}\right)^{3/2}
	\exp -\frac{2}{3}\frac{1}{\tau H_0} 
	\left(\frac{E}{\epsilon_0}\right)^{3/2}.
\end{equation}
The only case studied in the literature is a critical matter dominated
universe ($\Omega_0=8\pi G\rho_m/3H_0^2=1$, where $\rho_m$ is
the present matter density) with $m_j=0$ \cite{ReTu,KoTu,Biller}. In this
case and when $\tau \gg H_0^{-1}$ we obtain
\begin{equation} \label{previous}
\frac{d^2F_E}{dE\, d\Omega} = \frac{1}{4\pi} \frac{\tilde{n}_{\nu_i}(t_0)}
	{\tau H_0} \left(\frac{E}{m_i/2}\right)^{3/2},
\end{equation}
in agreement with previous results. 
  
For an open universe with matter density in critical density units
$\Omega_0 <1$ one has
\begin{eqnarray}
H(z) &=& H_0\sqrt{\Omega_k(1+z)^2+\Omega_0(1+z)^3}, \\
H_0 t(z) &=& \frac{1}{\Omega_k} 
	\frac{\sqrt{\Omega_k+\Omega_0(1+z)}}{1+z} 
	-\frac{\Omega_0}{\Omega_k^{3/2}}
	\ln \frac{\sqrt{\Omega_k}+\sqrt{\Omega_k+\Omega_0(1+z)}}
	{\sqrt{\Omega_0(1+z)}}, \label{tmattpluscurv}
\end{eqnarray}
being $\Omega_k \equiv 1- \Omega_0$.

The last example that we solve is a flat universe filled with matter and
a nonvanishing cosmological constant $\Lambda$. The contents in matter 
is given by $\Omega_0$ and the contribution of the cosmological constant is
$\Omega_\Lambda \equiv \Lambda/3H_0^2
= 1-\Omega_0$. The expansion rate and the expansion time at redshift $z$
are given, respectively, by 
\begin{eqnarray}
H(z) &=& H_0\sqrt{\Omega_\Lambda+\Omega_0(1+z)^3}, \\
H_0 t(z) &=& \frac{2}{3\sqrt{\Omega_\Lambda}} \ln
	\frac{\sqrt{\Omega_\Lambda}+\sqrt{\Omega_\Lambda +\Omega_0(1+z)^3}}
	{\sqrt{\Omega_0(1+z)^3}}. \label{tmattpluslamb}
\end{eqnarray}

We plot in Figs.~\ref{H0tau100}, \ref{H0tau10}, and~\ref{H0tau01} the
photon spectrum (\ref{restspectr}) in the three cases listed above,
for three different values of $\tau$, $\tau H_0=10$, $1$, and~$0.1$,
respectively. For an open universe and a flat universe with a nonzero
cosmological constant we choose the values of the density parameters
that seem to be observationally favored: $\Omega_0=0.3$ and
$\Omega_\Lambda=0.7$ \cite{observ}.

For the first model, $\Omega_0=1$ and $\Omega_\Lambda=0$,
one can see that the spectrum has a maximum at
\begin{equation} \label{MDmax}
E_{max} = \left( \frac{3}{2}\tau H_0\right)^{2/3} \! \epsilon_0 
	= \left( \frac{\tau}{t_0}\right)^{2/3}\! \epsilon_0 
	= \frac{\epsilon_0}{1+z(\tau)},
\end{equation}
with $E_{max}$ in the kinematically allowed region, $E_{max} <
\epsilon_0$, as long as $\tau < t_0$. The value of the normalized
spectrum at the maximum is
$(d^2F_E/dEd\Omega)/(\tilde{n}_{\nu_i}(t_0)/4\pi)= 3/2e=0.5518\ldots$
independent of $\tau H_0$. As $\tau H_0$ decreases the shape of the
spectrum becomes narrower and peaked to lower values of $E$. This
reflects the fact that when the lifetime $\tau$ is smaller than the
age of the universe $t_0$, most neutrinos have already decayed, and
the energy of the photons produced in the decay has been redshifted to
low values of $E$. The differences in the spectrum for the different
cosmological models are easy to understand. Let us compare the critical
matter dominated model with the flat model with nonvanishing $\Lambda$
(the open model falls in--between). For $E\rightarrow 0$ the photons are
produced at large $z_0$, when the Hubble expansion factor for the
nonvanishing $\Lambda$ model is smaller than $H(z)$ in the critical matter
dominated model by a factor $\sqrt{1-\Omega_\Lambda}$, while the term
$\exp -t/\tau$ in Eq.~(\ref{restspectr}) becomes unity for both
models.  Therefore, the spectrum is higher when $\Lambda > 0$ for $E$
small. On the other hand, when $E\rightarrow \epsilon_0$, the photons
have just been produced, at the same expansion rate $H_0$ but with an
age for the $\Lambda$ model older than the age for the critical matter
dominated case. Consequently, the term $\exp -t/\tau$ makes the
spectrum for the critical matter dominated model higher than for the flat
nonzero $\Lambda$ model in the range of large $E$. The position of the
maximum in the spectrum when one has a nonvanishing $\Lambda$ is given by
\begin{equation} \label{Lambmax}
E_{max} = \left( \frac{3}{2}\tau H_0\right)^{2/3} \left(
	\frac{2\Omega_0}{1+\sqrt{1+9\Omega_\Lambda \tau^2 H_0^2}}
	\right)^{2/3} \epsilon_0 < \left( \frac{3}{2}\tau H_0\right)^{2/3}
	\epsilon_0,
\end{equation}
which is smaller than the value of $E_{max}$ when~$\Lambda =0$.

We have shown that the differences in the shape of the photon spectrum
introduced by the cosmological energy density parameters are
small. Only a very accurate determination of the photon spectrum by
observation could be useful to discriminate among different
cosmological models, provided that neutrinos turned out to be unstable
with a priori known $\tau H_0\sim 1$. On the other hand, if the
photon spectrum was well measured and $\tau$ was unknown, the small
sensitivity of the spectrum to the energy contents of the universe
would allow a fair determination of $\tau$.

\section{Unstable neutrinos with a distribution of momenta}
\label{moving}

In this section we shall calculate the spectrum of photons produced by a
cosmic background of unstable neutrinos which are not at rest but decay
with a statistical distribution of momenta $f_i(\vec{p}_i)$. Besides
the expansion and geometric effects discussed in Sec.~\ref{rest}, the 
photon spectrum will also include features arising from this momentum 
distribution.

As always we consider the two--body decay 
\begin{equation} \label{twobody}
\nu_i (\epsilon_i, \vec{p}_i) \longrightarrow \nu_j (\epsilon_j, \vec{p}_j)
				+ \gamma (\epsilon,\vec{p}),
\end{equation}
where the energy and three--momentum of each particle are indicated.
A photon is produced with an energy $\epsilon$ at time $t(z)$; at
present its energy has been redshifted to $E$ with $\epsilon = (1+z)E$.

The present photon energy flux per solid angle with energy between $E$
and~$E+dE$ can be written as
\begin{equation}
\frac{d^2F_E}{d\Omega} =\frac{1}{4\pi}\int \frac{d^3\vec{p}}{(2\pi)^3}
	\frac{1}{(1+z)^3}\, E\, \delta f_\gamma(\vec{p},z),
\end{equation}
where $d^3\vec{p}/(2\pi)^3\: \delta f_\gamma(\vec{p},z)$ is the photon
number density, including both photon helicities, produced by
neutrinos decaying at redshift between $z$ 
and $z+dz$; these photons are produced with energy between $\epsilon$ and 
$\epsilon +d\epsilon$. The origin of the term $1/(1+z)^3$ is the
same as when the $\nu_i$'s decay at rest. The increment $\delta f_\gamma$
is given by the collision term of the Boltzmann equation \cite{Bern}
\begin{eqnarray}
\delta f_\gamma (\vec{p},z) = \frac{1}{2\epsilon}
	\int \frac{d^3\vec{p_i}}{(2\pi)^3 2\epsilon_i}
	\int \frac{d^3\vec{p_j}}{(2\pi)^3 2\epsilon_j}
	& &\left( 2\pi \right)^4 \delta^{(4)} 
	\left( p_i-p_j-p \right) |{\cal M}|^2 \times \nonumber \\ 
	& &\left[ f_i(1-f_j)(1+f_\gamma)-(1-f_i)f_jf_\gamma\right]
	\delta t. \label{Boltzmann}
\end{eqnarray}
The functions $f_j$ and $f_\gamma$ are the statistical momentum 
distributions of the decay products $\nu_j$ and $\gamma$, respectively. The
averaged square probability amplitude of the decay~(\ref{twobody})
is given~by
\begin{equation} \label{tau}
|{\cal M}|^2= 16\pi \frac{m_i^3}{\Delta m^2} \frac{1}{\tau}.
\end{equation}
For a fixed redshift $z$ (or fixed $\epsilon$) the time increment is
$\delta t = H(z)^{-1}dE/E$. The inverse process $\gamma \nu_j
\rightarrow \nu_i$ can be neglected when $m_i \gg m_j$ since at the
late times ($\tau \sim H_0^{-1}$) we are interested in the photon
energy is about $10^{-4}$ eV $\sim 3$ K, much smaller than the neutrino 
mass $m_i \sim 0.1$ eV. In order to obtain analytic expressions for the
final spectrum we will also neglect Pauli blocking and stimulated
emission, which are expected to produce small corrections. With these
approximations we can write
\begin{equation} \label{noquantumstat}
\frac{d^2F_E}{dE\, d\Omega} =\frac{1}{4\pi}\int 
	\frac{d^3\vec{p}}{(2\pi)^3 2 \epsilon} 
	\int \frac{d^3\vec{p_i}}{(2\pi)^3 2\epsilon_i}
	\int \frac{d^3\vec{p_j}}{(2\pi)^3 2\epsilon_j}
	\left( 2\pi \right)^4 \delta^{(4)} 
	\left( p_i-p_j-p \right) |{\cal M}|^2 
	\frac{1}{H(z)}\frac{f_i}{(1+z)^3}.
\end{equation}

We shall now relate $f_i$ at time $t(z)$ with the distribution function
$\tilde{f}_i$ that the neutrinos $\nu_i$ would have at present if they 
had not decayed. The evolution of $f_i$ is governed by the 
Boltzmann~equation
\begin{eqnarray}
\frac{df_i}{dt}=-\frac{1}{2\epsilon_i}
	\int \frac{d^3\vec{p_j}}{(2\pi)^3 2\epsilon_j}
	\int \frac{d^3\vec{p}}{(2\pi)^3 2\epsilon}
	& &\left( 2\pi \right)^4 \delta^{(4)} 
	\left( p_i-p_j-p \right) |{\cal M}|^2 \times \nonumber \\ 
	& &\left[ f_i(1-f_j)(1+f_\gamma)-(1-f_i)f_jf_\gamma\right] .
\end{eqnarray} 
Neglecting Pauli blocking, Bose stimulation, and the inverse process, and
using Eq.~(\ref{tau}) we get the differential equation
\begin{equation}
\frac{df_i(\vec{p}_i(t),t) }{dt} = 
	-\frac{1}{\tau} \frac{m_i}{\epsilon_i} f_i(\vec{p}_i(t),t),
\end{equation}
which can easily be solved to obtain
\begin{equation}
f_i(\vec{p}_i(t),t) = f_i(\vec{p}_i(t_p),t_p) \exp -\int_{t_p}^t \!
		\frac{dt'}{\tau} \frac{m_i}{\epsilon_i(t')},
\end{equation}
where we choose the arbitrary constant time $t_p$ to be 
$t_p \ll \tau $. Including the $(1+z)^{-3}$ that appears
in~(\ref{noquantumstat}) we can write
\begin{eqnarray}
\frac{f_i(\vec{p}_i(t),t)}{(1+z)^3} d^3\vec{p}_i(t) &=& 
	\frac{f_i(\vec{p}_i(t_p),t_p)}{(1+z(t_p))^3} d^3\vec{p}_i(t_p) 
	\exp -\int_{t_p}^t \! \frac{dt'}{\tau} 
	\frac{m_i}{\epsilon_i(t')} \nonumber \\
	&=& \tilde{f}_i(\vec{q},t_0) d^3\vec{q} 
	\, \exp -\int_{t_p}^t \!
	\frac{dt'}{\tau} \frac{m_i}{\epsilon_i(t')}, \label{decaylaw}
\end{eqnarray} 
where $\vec{q}\equiv \vec{p}_i(t_p)/(1+z(t_p))=\vec{p}_i(t)/(1+z)$
is the neutrino momentum at $t_0$. The distribution function 
$\tilde{f}_i(\vec{q},t_0)$ would be the neutrino distribution function
at present if the $\nu_i$'s were stable. In appendix~\ref{distrib}
we derive its expression. 

Making use of Eqs.~(\ref{tau}), (\ref{noquantumstat}), and~(\ref{decaylaw})
we get
\begin{eqnarray}
\frac{d^2F_E}{dE\, d\Omega} = \frac{1}{4\pi} & & 
	\int \frac{d^3\vec{p}}{(2\pi)^3 2 \epsilon} 
	\int \frac{d^3\vec{q}}{(2\pi)^3 2\epsilon_i}
	\int \frac{d^3\vec{p_j}}{(2\pi)^3 2\epsilon_j} 
	\left( 2\pi \right)^4 \delta^{(3)} 
	\left( \vec{p}_i-\vec{p}_j-\vec{p} \right)\times \nonumber \\
	& &\delta \left( \epsilon_i-\epsilon_j-\epsilon \right)
	16\pi \frac{m_i^3}{\Delta m^2} \frac{1}{\tau}
	\frac{1}{H(z)}
	\tilde{f}_i(\vec{q},t_0)
	\, \exp -\int_{t_p}^t \!
	\frac{dt'}{\tau} \frac{m_i}{\epsilon_i(t')},
\end{eqnarray}
with
\begin{eqnarray}
\epsilon_{i,j} &=& \sqrt{m_{i,j}^2+\vec{p}_{i,j}^{\, 2}},\\
\vec{p}_i &=& (1+z) \vec{q}, \\
\epsilon &=& |\vec{p}| = (1+z) E. \label{epsilonz}
\end{eqnarray}
The momentum $\vec{p}_j$ can be integrated out using 
$\delta^{(3)}(\vec{p}_i-\vec{p}_j-\vec{p})$. The lower limit of the
integral in the exponential can be set to 0, $t_p=0$. Assuming
isotropy for $\tilde{f_i}$ and using Eq.~(\ref{epsilonz})
we can finally write the following expression for the photon spectrum 
produced by the decaying neutrinos 
\begin{equation} \label{movingspectr}
\frac{d^2F_E}{dE\, d\Omega} = \int_0^{t_0} \! dt \; I(z(t)) 
	= \int_0^\infty \! \frac{dz}{1+z} \frac{1}{H(z)}I(z), 
\end{equation}
where $I(z(t))$ is the energy flux per unit photon energy and solid angle
produced between $t$ and $t+dt$ and is given by
\begin{eqnarray}
I(z) &\equiv& \frac{1}{4\pi} \int \! \frac{d^3 \vec{q}}{(2\pi)^3}
	\; \tilde{f}_i(\vec{q}) \; \frac{1}{\tau} \frac{m_i}{\epsilon_i} \;
	\exp -\frac{{\cal T}(y,z)}{\tau} \nonumber \\ 
	& & \times (1+z)E \; \left(\frac{1+z}{1+z_0}\right)^2 
	\delta \left( (1+z)E - 
	\frac{\Delta m^2/2}{\epsilon_i-|\vec{p}_i| \cos \theta}\right).
	\label{Iz}
\end{eqnarray}
The redshift $z_0$ has been defined in Eq.~(\ref{zzero}). The angular
variable $\theta$ is the angle between the momentum of 
the decaying neutrino and the momentum of the emitted photon, and
\begin{equation}
y\equiv \frac{|\vec{q}|}{m_i}.
\end{equation}
We have defined the ``kinematically contracted expansion time''
\begin{equation} \label{integral}
{\cal T} (y,z)  \equiv \int_0^{t(z)}\! dt\; \frac{m_i}{\epsilon_i(z(t))} =
	\int_z^\infty \! \frac{dz'}{1+z'} 
	\frac{1}{H(z')} \frac{1}{\sqrt{1+y^2(1+z')^2}},
\end{equation}
which reduces to the ordinary expansion time only if $y=0$,
${\cal T}(0,z)=t(z)$. In appendix~\ref{contract} we calculate
analytically the function ${\cal T} (y,z)$ for the three scenarios
defined in Sec.~\ref{rest}.

Equation~(\ref{movingspectr}) has to be multiplied by a factor $2$ if,
as is usually assumed, there is a cosmological population of 
antineutrinos $\bar{\nu}_i$ with the same distribution function
as the $\nu_i$ and the CP symmetry is not broken in the decay.

The physical meaning of the different terms in Eq.~(\ref{Iz}) is
straightforward: there is a sum over all possible momenta with which
the neutrinos $\nu_i$ can decay, each particular momentum weighed by
the decay probability per unit time
$\tau^{-1}(m_i/\epsilon_i)\tilde{f}_i\exp-{\cal T}/\tau$. The time
dilation factor $\epsilon_i/m_i$ that appears multiplying $\tau$, in
the factor before the exponential  and in the exponent, takes into
account the relativistic delay in the decay of moving neutrinos. The term
$(1+z)$ accompanying $E$ stems from the cosmic expansion, which
redshifts the photon energy. The term $\left((1+z)/(1+z_0)\right)^2$
accounts for the photon phase space. Finally, Dirac's delta  enforces
energy conservation. Each photon is Doppler shifted by the motion of
the neutrino that emits it. The energy of the emitted photon in the
rest frame of the decaying neutrino $\epsilon_0$ is blueshifted if
$\cos \theta > 0$ (emitter $\nu_i$ and emitted $\gamma$ moving in the
same direction) or redshifted if $\cos \theta < 0$ ($\nu_i$ and
$\gamma$ moving in opposite directions).

Equations~(\ref{movingspectr}) and~(\ref{Iz}) are the main result of
this section, they give the spectrum of photons produced by the decay
of a cosmic background of neutrinos. They hold for an arbitrary
Robertson--Walker geometry and equation of state of the universe, and
for any isotropic, nondegenerate momentum distribution of the
neutrino background.

As a consistency check we evaluate Eqs.~(\ref{movingspectr})
and~(\ref{Iz}) when $\tilde{f}_i(\vec{q}) =
(2\pi)^3  \tilde{n}_{\nu_i}(t_0) \delta^{(3)}(\vec{q})$,
i. e., all the $\nu_i$ are at rest, and we
recover~Eq.~(\ref{restspectr}).

We can also calculate the total photon flux per solid angle at present
time produced by unstable neutrinos with an arbitrary
$\tilde{f}_i(\vec{q})$
\begin{equation}
\frac{dF_n}{d\Omega} \equiv \int \! \frac{dE}{E}\; 
	\frac{d^2F_E}{dE\; d\Omega} =
	\frac{1}{4\pi} \int \! \frac{d^3 \vec{q}}{(2\pi)^3} 
	\; \tilde{f}_i(\vec{q})\;
	\left( 1-\exp -\frac{{\cal T} (y,0)}{\tau} \right).
\end{equation}
If $\tau \ll H_0^{-1}$ we find 
\begin{equation}
\frac{dF_n}{d\Omega} = \frac{1}{4\pi} \int \! 
	\frac{d^3 \vec{q}}{(2\pi)^3} \; \tilde{f}_i(\vec{q}) = 
	\frac{1}{4\pi} \, \tilde{n}_{\nu_i}(t_0),
\end{equation}
as one should expect, since in this regime all $\nu_i$ have already
decayed into photons.

Next we shall consider the example in which $\tilde{f}_i$ is given by the
momentum distribution of the standard stable neutrino background (see
appendix~\ref{distrib})
\begin{equation} \label{backgr}
\tilde{f}_i (\vec{q}) = \frac{1}{\exp |\vec{q}|/T_0 \; +1},
\end{equation}
where $T_0$ is the present neutrino temperature and we have set the neutrino
chemical potential to zero. We shall study the limit $m_i \gg T_0$.
The motion of the nonrelativistic $\nu_i$ will introduce small 
corrections to the photon spectrum when compared to the spectrum produced 
by neutrinos decaying at rest. In order to calculate these
corrections we shall need the asymptotic expressions of ${\cal T} (y,z)$
when $y \equiv |\vec{q}|/m_i \rightarrow 0$, which are calculated
in appendix~\ref{contract}. In most cases we can write
\begin{equation} \label{power}
H_0 {\cal T} (y,z) = H_0 t(z) - \alpha y^n + \ldots
\end{equation}
where $\alpha$ is a constant and we neglect terms that decrease 
faster than $y^n$ when $y \rightarrow 0$. Plugging Eqs.~(\ref{power})
and~(\ref{backgr})
into Eqs.~(\ref{movingspectr}) and~(\ref{Iz}) we obtain
\begin{equation} \label{powerspect}
\frac{d^2F_E}{dE\; d\Omega} = \frac{1}{4\pi} \frac{\tilde{n}_{\nu_i}(t_0)}
	{\tau H(z_0)} \exp -\frac{t(z_0)}{\tau} 
	\left( 1+ 
	\frac{(2-\frac{1}{2^{n+1}})\zeta (n+3)\Gamma (n+3)\alpha}
	{3\zeta (3)\tau H_0} \left( \frac{T_0}{m_i} \right)^n
	+ \ldots \right),
\end{equation}
where $\zeta (x)$ is the Riemann zeta function and $\Gamma (x)$ is the
Euler gamma function. For instance, in a universe with 
$\Omega_0\sim 1$ one has $n=3/2$ and $\alpha = \Omega_0^{-1/2} 
\Gamma (1/4)^2/6\sqrt{\pi}$, see Eqs.~(\ref{ymatt}), (\ref{ymattcurv}), 
and~(\ref{ymattlamb}). Equation~(\ref{powerspect}) is then
\begin{equation}
\frac{d^2F_E}{dE\; d\Omega} = \frac{1}{4\pi} \frac{\tilde{n}_{\nu_i}(t_0)}
	{\tau H(z_0)} \exp -\frac{t(z_0)}{\tau} \left( 1+
	\frac{35(1-\frac{1}{8\sqrt{2}})\zeta (\frac{9}{2})
	\Gamma \left(\frac{1}{4}\right)^2}{48\zeta (3)
	\Omega_0^{1/2}\tau H_0} \left( \frac{T_0}{m_i}
	\right)^{3/2} + \ldots \right),
\end{equation}
with $\zeta (9/2)=1.0547\ldots$, $\zeta (3)=1.2021\ldots$,
and~$\Gamma (1/4)=3.6256\ldots$

It is also possible to obtain the following asymptotic 
form of ${\cal T} (y,z)$
\begin{equation} \label{log}
H_0 {\cal T} (y,z) = H_0 t(z) + \beta y^n \ln y + \ldots
\end{equation}
with $\beta$ a constant, which renders
\begin{equation} \label{logspect}
\frac{d^2F_E}{dE\; d\Omega} = \frac{1}{4\pi} \frac{\tilde{n}_{\nu_i}(t_0)}
	{\tau H(z_0)} \exp -\frac{t(z_0)}{\tau} 
	\left( 1+ 
	\frac{(4-\frac{1}{2^n})\zeta (n+3)\Gamma (n+3) \beta}
	{3\zeta (3) \tau H_0} \left( \frac{T_0}{m_i} \right)^n
	\ln \frac{m_i}{T_0} + \ldots \right).
\end{equation}
For example, in a critical radiation dominated universe we have to make
use of Eq.~(\ref{logspect}) with $n=2$ and $\beta = 1/2$,
see Eq.~(\ref{yrad}),
\begin{equation}
\frac{d^2F_E}{dE\; d\Omega} = \frac{1}{4\pi} \frac{\tilde{n}_{\nu_i}(t_0)}
	{\tau H(z_0)} \exp -\frac{t(z_0)}{\tau} \left( 1+ 
	\frac{15\zeta (5)}{2\zeta (3)\tau H_0}
	\left( \frac{T_0}{m_i}\right)^2
	\ln \frac{m_i}{T_0} + \ldots \right),
\end{equation}
with $\zeta (5)=1.0369\ldots$

In all the above examples the leading correction is independent
of $E$ and positive. There
are more photons at present, with a given energy $E$, compared to the
case when the  neutrinos decay at rest. The dominant effect is the
time dilation in the decay of the neutrinos included in ${\cal T}
(y,z)/\tau$. Photons with present energy $E$ are produced at $1+z \sim
\epsilon_0/E$; there are more neutrinos available at time $t(z)$ if
the decay is delayed, and therefore  more photons with energy $E$ can
be produced. There is a second effect of time delay that goes in
opposite direction, the number of photons produced at $t(z)$ decreases
since the decay probability is inversely proportional to the dilation
factor, but this second effect is order $(T_0/m_i)^2$ and therefore
subleading.

We notice that, in general, the natural expansion parameter for ${\cal
T} (y,z)$ is $y^{1/2}$ or equivalently $(T_0/m_i)^{1/2}$ in the final
photon spectrum, if we assume the distribution~(\ref{backgr}). High
precision determination of the photon spectrum would require analytic
calculation of higher orders in $(T_0/m_i)^{1/2}$. In addition, the
small corrections introduced by Pauli blocking and stimulated emission
can be important when compared with the relativistic corrections we
have calculated \cite{refpapers}. An accurate determination of the
photon spectrum including all these statistical effects can only be
performed numerically, which goes beyond the scope of the present
paper. We think that the analytic expressions that we have found can be
useful to understand the physics of a late decaying cosmic
neutrino background, in spite of the fact that we neglect
contributions that should be included in a complete numerical study.

While in the case that the neutrinos decay at rest, the photon flux
vanishes when $E > \epsilon_0$, now it is possible to have a photon flux
with energy higher than $\epsilon_0$, albeit very small, since when a
moving neutrino decays it can emit a Doppler blueshifted photon
with energy higher than $\epsilon_0$. From Eqs.~(\ref{movingspectr}),
(\ref{Iz}), and~(\ref{backgr}) we obtain
\begin{equation}
\frac{d^2F_E}{dE\; d\Omega} = \frac{1}{4\pi} \frac{\tilde{n}_{\nu_i}(t_0)}
	{\tau H_0} \frac{1}{3\zeta (3)} \frac{m_i}{T_0}
	\left( \frac{E}{\epsilon_0}\right)^2
	\exp -\frac{1}{2} \frac{m_i}{T_0}\frac{E}{\epsilon_0}, 
\end{equation}
valid for $E\gg \epsilon_0$, $m_i \gg T_0$,  and $\tau H_0$ not too
small (so that the term $\exp -{\cal T}/\tau$ does not dominate the
fall of the spectrum).

To end this section we would like to point out that the  photon
spectrum obtained is not a thermal spectrum even if the neutrinos,
that act as the photon source, are distributed thermally. The cosmic
expansion and the decay kinematics render a photon distribution that
is out of equilibrium. Later interaction of the decay products with
matter could restore thermal equilibrium. However, we always assume
that the decay happens after decoupling of matter and radiation, i. e.,
$\tau \gg 300,000$ yr, therefore the decay produced photons never attain
thermal equilibrium.

\section{Conclusions} \label{concl}

In this paper we have studied the contribution to the photon
background of a cosmological density of decaying neutrinos. 

We first have adopted the assumption that neutrinos decay at rest and
have investigated the fluxes in different cosmological scenarios: 1) a
power--law cosmic expansion, 2) an open universe filled only with
matter, and finally 3) a flat universe with matter and a nonzero
cosmological constant $\Lambda$. We have calculated the exact formulae
for these cases, and found the maximum of the spectrum as a function
of the photon energy. 

The differences amongst the scenarios are not large. This result has
the consequence that, when the observed photon spectrum is used to
constrain the neutrino lifetime, the limit does not depend too much on
the energy content of the universe. 

Second, we have studied the consequences for the photon background
when the decaying neutrinos have a momentum distribution. We have
worked out the relevant formulae for the scenarios 1), 2), and 3)
mentioned earlier. We have also particularized our equations to the
standard neutrino background. In all the cases, to compare with the
photon spectrum calculated for neutrinos at rest, we have determined
the first relativistic correction in our formulae. 

In the literature only the case of neutrinos decaying at rest in a
critical matter dominated universe has been treated; this corresponds
to Eq.~(\ref{previous}). In this paper we have performed a general
study of the photon spectrum that will allow to obtain reliable bounds
on electromagnetic neutrino properties using photon background
observations. Also, in the case of an eventual positive signal, the
spectrum we have calculated would be helpful to obtain an
observational determination of the lifetime $\tau$.

A final comment is that we believe that our calculations may be useful
when considering scenarios of radiatively decaying particles. For
example, of the type considered by Sciama \cite{Sciama} that might be
the explanation for the observed ionization of intergalactic
hydrogen. As another example, our results can be applied to the
contribution to the photon background coming from a relic scalar or
pseudoscalar particle density decaying into two photons of the type
considered in \cite{phi1,phi2}. Of course one should introduce the
appropriate changes in the equations of our paper.

\acknowledgments

This work was partially supported by the CICYT Research Project AEN98--1116
at IFAE/UAB and by the DOE and the NASA grant NAG 5--7092 at Fermilab.
R.T. is financially supported by a postdoctoral fellowship from the~MEC.

\appendix

\section{Distribution of momenta for a stable neutrino
cosmic background} \label{distrib}

In this appendix we shall study the momentum distribution of a relic
cosmic background of stable neutrinos with mass $m$. We
neglect the neutrino chemical potential setting it to zero although
our results could easily be generalized to include a nonvanishing
chemical potential.

Neutrinos decouple from the primordial plasma of photons, electrons
and baryons at $t_d\sim 1$ sec., when the temperature is $T_d \sim 1$
MeV. Hereafter they freely stream without any further
nongravitational interactions. Particle number conservation gives the
following relation between the momentum distribution at time $t>t_d$ and
the momentum distribution at decoupling time $t_d$
\begin{equation}
f(\vec{k},t) = f(\vec{k}_d,t_d), 
\end{equation}
with the neutrino momenta related by $\vec{k}=\vec{k}_d\: a_d/a$,
being $a$ the cosmic expansion factor. Until time $t_d$ the distribution
function is given by the Fermi--Dirac thermal distribution, therefore
\begin{equation}
f(\vec{k},t) = \frac{1}{\exp E_d/T_d \; +1}.
\end{equation} 
The energy at decoupling $E_d = \sqrt{m^2+\vec{k}^2_d}$ can be written as
\begin{equation} 
\frac{E_d}{T_d} = \sqrt{\frac{m^2}{T_d^2}+\frac{\vec{k}^2}{T^2}},
\end{equation}
where the neutrino temperature $T$ is related to the neutrino temperature
at decoupling $T=T_d\: a_d/a$. Writing everything together we obtain
\begin{equation} \label{distfunc}
f(\vec{k},t) = \left(1+\exp \sqrt{\frac{m^2}{T_d^2}+
	\frac{\vec{k}^2}{T^2}}\right)^{-1}.
\end{equation}
This distribution reduces to a thermal distribution only if the
neutrinos are extremely relativistic, $T\gg m$, or nonrelativistic,
$T\ll m$, from decoupling time until present (for nonrelativistic
neutrinos one has to redefine the temperature)~\cite{KoTu}. The
neutrino temperature $T$ is related to the photon temperature by the
relation $T=(4/11)^{1/3} T_\gamma$. We are interested in a neutrino
mass about $m \sim 0.1$ eV, hence the neutrinos are extremely
relativistic at decoupling time but nonrelativistic at present because
now $T_0\sim 1.7 \times 10^{-4}$ eV. Neglecting the term $m/T_d \ll 1$
in Eq.~(\ref{distfunc}) we finally obtain
\begin{equation} \label{finaldist}
f(\vec{k},t) = \frac{1}{\exp |\vec{k}|/T(t) \; +1}.
\end{equation}
Note that, in general, this is not a thermal distribution since in
the Fermi--Dirac distribution one has $E=\sqrt{m^2+\vec{k}^2}$ which in
our case has been replaced by $|\vec{k}|$. Strictly speaking we cannot
call the parameter $T$ neutrino temperature, but we use this term
anyway since it does not cause any confusion. We make use of
Eq.~(\ref{finaldist}) in Sec.~\ref{moving} to calculate the spectrum
of decay produced photons.

\section{Calculation of the function ${\cal T}$} \label{contract}

The integral ${\cal T} (y,z)$ defined in Eq.~(\ref{integral}) 
can be solved in terms of elementary and special functions. We 
shall consider separately the three 
cases introduced in Sec.~\ref{rest}. 

\subsection{Power--law cosmic expansion}

When the Hubble function is given by the power--law (\ref{Hpower}) we
have to calculate the following integral
\begin{equation} \label{powerlaw}
H_0 {\cal T} (y,z) = \int_{1+z}^\infty \! \frac{dv}{v^{n+1}} 
	\frac{1}{\sqrt{1+y^2v^2}},
\end{equation}
where $n\geq 0$.
This integral can be related, by means of the 
change of integration variable $u=y^2v^2/(1+y^2v^2)$, to the incomplete 
beta function, which in turn is related to the hypergeometric function
$F(\alpha, \beta; \gamma ; u) =\;  _2F_1(\alpha, \beta; \gamma ; u)$
\cite{GrRy}.

For $n$ not an even integer we obtain:
\begin{eqnarray} 
H_0 {\cal T} (y,z)  &=& \frac{\Gamma(-\frac{n}{2})
	\Gamma(\frac{n+1}{2})}{2\sqrt{\pi}} y^n \nonumber \\ 
	& & + \frac{1}{n} \left( \frac{\sqrt{1+y^2(1+z)^2}}{1+z} \right)^n
	F\left( -\frac{n}{2},\frac{1-n}{2}; 1-\frac{n}{2}; 
	\frac{y^2(1+z)^2}{1+y^2(1+z)^2}\right). \label{notevenint}
\end{eqnarray}

For $n$ an even integer, $n>0$, the solution can always be given in terms of 
elementary functions
\begin{eqnarray} 
H_0 {\cal T} (y,z)  &=& (-1)^{n/2} \frac{n!}{(n!!)^2}
	y^n \ln \frac{\sqrt{1+y^2(1+z)^2}+1}{y(1+z)}
	+\frac{\sqrt{1+y^2(1+z)^2}}{(1+z)^n} \times \nonumber \\
	& & \left( \frac{1}{n}-\frac{n-1}{n(n-2)}y^2(1+z)^2+ \ldots 
	(-1)^{1+n/2} \frac{n!}{(n!!)^2} 
	y^{n-2}(1+z)^{n-2} \right). \label{evenint}
\end{eqnarray}

Both Eqs.~(\ref{notevenint}) 
and~(\ref{evenint}) can be written  
in a more compact but less insightful way using the 
transformation formulae of the hypergeometric function \cite{GrRy}
\begin{equation} \label{compact}
H_0 {\cal T} (y,z) = \frac{1}{n+1} \frac{1}{\sqrt{1+y^2(1+z)^2}} 
	\frac{1}{(1+z)^n}
	F\left( 1,\frac{1}{2};\frac{n+3}{2};\frac{1}{1+y^2(1+z)^2}\right),
\end{equation}
which holds for any $n\geq 0$.

As particular examples of Eq.~(\ref{notevenint}), we write down 
${\cal T} (y,z)$ for a curvature and a flat matter dominated
universe, $n=1,\; 3/2$, respectively,
\begin{eqnarray}
H_0 {\cal T} (y,z)|_{n=1}    &=& -y + \frac{\sqrt{1+y^2(1+z)^2}}{1+z}, \\
H_0 {\cal T} (y,z)|_{n=3/2}  &=& -\frac{1}{6\sqrt{\pi}}\Gamma (1/4)^2y^{3/2} 
	\nonumber \\
	& & +\frac{2}{3}\frac{\left( 1+y^2(1+z)^2\right)^{3/4}}{(1+z)^{3/2}}
	F\left( -\frac{3}{4},-\frac{1}{4};\frac{1}{4};
	\frac{y^2(1+z)^2}{1+y^2(1+z)^2} \right) \\
	&=& -\frac{1}{6\sqrt{\pi}}\Gamma (1/4)^2y^{3/2} 
	+\frac{2}{3} \frac{\sqrt{1+y^2(1+z)^2}}{(1+z)^{3/2}} \nonumber \\
	& & + \frac{1}{3}y^{3/2}
	F\left( \arccos \frac{1-y^2(1+z)^2}{\left( 1+y(1+z)\right)^2},
	\frac{1}{\sqrt{2}}  \right).
\end{eqnarray}
The function $F(\phi,k)$ is the incomplete elliptic integral of the
first kind, see Sec.~\ref{openuniv}.
In the limit $y\rightarrow 0$ the above expressions become
\begin{eqnarray}
H_0 {\cal T} (y,z)|_{n=1}    &=& H_0t(z)|_{n=1} -y + O(y^2),\\
H_0 {\cal T} (y,z)|_{n=3/2}  &=& H_0t(z)|_{n=3/2} - \frac{1}{6\sqrt{\pi}}
		\Gamma (1/4)^2y^{3/2} + O(y^2), \label{ymatt}
\end{eqnarray}
where $t(z)|_n$ is given by Eq.~(\ref{tpowerlaw}) and
$\Gamma (1/4)^2/6\sqrt{\pi} = 1.2360\ldots$

As an example of Eq.~(\ref{evenint}) one has a critical radiation
dominated universe, $n=2$,
\begin{equation}
H_0 {\cal T} (y,z)|_{n=2} = -\frac{y^2}{2} 
	\ln \frac{1+\sqrt{1+y^2(1+z)^2}}{y(1+z)}
	+\frac{1}{2}\frac{\sqrt{1+y^2(1+z)^2}}{(1+z)^2}.
\end{equation}
In the limit $y\rightarrow 0$ we find
\begin{equation} \label{yrad}
H_0 {\cal T} (y,z)|_{n=2} = H_0t(z)|_{n=2} + \frac{1}{2} y^2 \ln y + O(y^2),
\end{equation}
with $t(z)|_{n=2}$ given by Eq.~(\ref{tpowerlaw}).

As a final example of power law we consider a critical $\Lambda$ dominated 
universe, $n=0$. Directly integrating 
(\ref{powerlaw}), or using (\ref{compact}) (see representation
of elementary functions in terms of a hypergeometric function
in \cite{GrRy}) we obtain
\begin{equation}
H_0 {\cal T} (y,z)|_{n=0} = \ln \frac{1+\sqrt{1+y^2(1+z)^2}}{y(1+z)}.
\end{equation}
When $y\rightarrow 0$ this expression diverges because $t(z)$ diverges
for a critical $\Lambda$ dominated universe.

\subsection{Open universe} \label{openuniv}

The integral to solve now is
\begin{equation} \label{mattpluscurv}
H_0 {\cal T} (y,z) = \int_{1+z}^\infty \! \frac{dv}{v} 
	\frac{1}{\sqrt{\Omega_k v^2 +
	\Omega_0 v^3}} \frac{1}{\sqrt{1+y^2v^2}}.
\end{equation}
This integral can be written as
\begin{equation}
\int_{1+z}^\infty \! \frac{dv}{v^2} \frac{1}{\sqrt{P_3(v)}},
\end{equation}
with $P_3(v)$ a cubic polynomial, which is a particular case of elliptic 
integral \cite{ByFr}. By means of the variable 
change cn $\! u =$ cn $\! (u,k) = (v+a-A)/(v+a+A)$, where
cn $\! u$ is the Jacobian cosine amplitude $u$, $a\equiv \Omega_k/\Omega_0$,
$A^2 \equiv a^2+1/y^2$, and $k^2 \equiv (A+a)/2A$, the elliptic 
integral~(\ref{mattpluscurv}) can be written
linearly in terms of elementary functions and the three fundamental 
incomplete elliptic integrals, of the first kind~$F$, second kind~$E$, and 
third kind~$\Pi$ \cite{ByFr,error}:  
\begin{eqnarray}
H_0 {\cal T} (y,z)  &=& \frac{y^{1/2}}{\Omega_k}
	\left( \Omega_0^2+\Omega_k^2 y^2\right) ^{1/4}
        \left( 1-\frac{\Omega_ky}{\sqrt{\Omega_0^2+\Omega_k^2 y^2}}
	\right) F(\phi,k) \nonumber \\ 
	& & -\frac{y^{1/2}}{\Omega_k}
	\left( \Omega_0^2+\Omega_k^2 y^2\right) ^{1/4}
	E(\phi,k) \nonumber \\
	& & +\frac{1}{4\Omega_k^2y^{1/2}}\frac{\left( 
	\sqrt{\Omega_0^2 + \Omega_k^2 y^2}-\Omega_ky\right) ^2}{\left( 
	\Omega_0^2+\Omega_k^2 y^2\right) ^{1/4}} 
	\Pi (\phi,n^2,k) \nonumber\\
	& & -\frac{\Omega_0}{4\Omega_k^{3/2}} \ln \left| 
	\frac{\sqrt{\Omega_k}\sqrt{1+y^2(1+z)^2}+
	\sqrt{\Omega_k+\Omega_0(1+z)}}
	{\sqrt{\Omega_k}\sqrt{1+y^2(1+z)^2}-
	\sqrt{\Omega_k+\Omega_0(1+z)}}\right| \nonumber\\
	& & +\frac{1}{\Omega_k}\frac{\left( \sqrt{\Omega_0^2+\Omega_k^2y^2}
	+\Omega_ky\right) 
	\sqrt{\Omega_k+\Omega_0(1+z)}}{\sqrt{\Omega_0^2+\Omega_k^2y^2} +
	y\left( \Omega_k+\Omega_0(1+z)\right)} 
	\frac{\sqrt{1+y^2(1+z)^2}}{1+z}. \label{elliptics}
\end{eqnarray}
We use the following definitions of the incomplete elliptic 
integrals \cite{ByFr}
\begin{eqnarray}
F(\phi,k) &\equiv& \int_0^\phi \! \frac{d\theta}{\sqrt{1-k^2\sin^2\theta}},\\
E(\phi,k) &\equiv& \int_0^\phi \! d\theta \; \sqrt{1-k^2\sin^2\theta},\\
\Pi (\phi,n^2,k) &\equiv& \int_0^\phi \! 
\frac{d\theta}{(1-n^2\sin^2\theta)\sqrt{1-k^2\sin^2\theta}}.
\end{eqnarray}
The variables $\phi$, $k$, and $n$ of the three elliptic integrals are 
related to the physical variables of our problem:
\begin{eqnarray}
\cos \phi &\equiv& \frac{\left( \Omega_k+\Omega_0(1+z) \right)y-
	\sqrt{\Omega_0^2+\Omega_k^2y^2}}{\left( \Omega_k+
	\Omega_0(1+z) \right)y+\sqrt{\Omega_0^2+\Omega_k^2y^2}}, \\
k^2 &\equiv& \frac{1}{2}\left( 1+
	\frac{\Omega_ky}{\sqrt{\Omega_0^2+\Omega_k^2y^2}}\right), \\
n^2 &\equiv& \frac{1}{4\Omega_ky}\sqrt{\Omega_0^2+\Omega_k^2y^2} \left( 1+
	\frac{\Omega_ky}{\sqrt{\Omega_0^2+\Omega_k^2y^2}}\right) ^2.
\end{eqnarray}

When $y\rightarrow 0$ and $\Omega_0$ finite Eq.~(\ref{elliptics})
reduces to \cite{note2}
\begin{equation} \label{ymattcurv}
H_0 {\cal T} (y,z) = H_0 t(z) - \frac{1}{6\sqrt{\pi}}\Gamma (1/4)^2
	\frac{1}{\Omega_0^{1/2}}y^{3/2} + O(y^2),
\end{equation}
where $t(z)$ is given by Eq.~(\ref{tmattpluscurv}).

For $\Omega_0 \rightarrow 0$ and $y$ finite Eq.~(\ref{elliptics}) gives
\begin{equation}
H_0 {\cal T} (y,z) = H_0 {\cal T} (y,z)|_{n=1} + 
	\frac{1}{2} \left( H_0 {\cal T} (y,z)|_{n=1}
	- \ln \frac{1+\sqrt{1+y^2(1+z)^2}}{y(1+z)}
	\right) \Omega_0 + \ldots
\end{equation}
where the ellipsis means smaller contributions like $\Omega_0^2$, 
$\Omega_0^2 \ln \Omega_0$, and so on. 

\subsection{Flat universe with nonzero $\Lambda$}

In the last case that we study we have to cope with the integral

\begin{equation} \label{mattpluslamb}
H_0 {\cal T} (y,z) = \int_{1+z}^\infty \! \frac{dv}{v} 
	\frac{1}{\sqrt{\Omega_\Lambda +\Omega_0 v^3}} 
	\frac{1}{\sqrt{1+y^2v^2}},
\end{equation}
which is of the class
\begin{equation}
\int_{1+z}^\infty \! \frac{dv}{v} \frac{1}{\sqrt{P_5(v)}},
\end{equation}
being $P_5(v)$ a five degree polynomial. 
Integrals of this sort can be solved as
linear combinations of elementary functions and hyperelliptic integrals
\cite{ByFr}. Now there are five fundamental incomplete hyperelliptic 
integrals, two of the first kind, two of the second kind and one 
of the third kind. Since the solution in terms of the hyperelliptic 
integrals would not be particularly enlightening, we only study two 
limiting cases of~(\ref{mattpluslamb}).

For $y\rightarrow 0$ and $\Omega_0$ finite we find
\begin{equation} \label{ymattlamb}
H_0 {\cal T} (y,z) = H_0 t(z) - \frac{1}{6\sqrt{\pi}}\Gamma (1/4)^2
	\frac{1}{\Omega_0^{1/2}}y^{3/2} + O(y^2),
\end{equation}
where $t(z)$ is given by Eq.~(\ref{tmattpluslamb}).
In the regime $\Omega_0 \rightarrow 0$ and $y$ finite we obtain
\begin{equation}
H_0 {\cal T} (y,z) = H_0 {\cal T} (y,z)|_{n=0} -\frac{b}{y}\Omega_0^{1/3} 
+ \ldots
\end{equation}
with
\begin{equation}
b \equiv 1+\frac{F\left( \frac{2}{3},\frac{4}{3};\frac{5}{3};
	\frac{1-\sqrt{2}}{2}\right)}
	{2^{4/3}(1+\sqrt{2})^{2/3}} = 1.199\ldots
\end{equation}


\begin{thebibliography}{99}
\bibitem{note} For a cosmic fluid with equation of state $p=w\rho$,
with $w$ constant, and
no spatial curvature one obtains Eq.~(\ref{Hpower}) with $n=3(1+w)/2$.
The relativistic condition $w\leq 1$ constrains $n$ to be $n \leq 3$.
\bibitem{error} Formula 361.54 in \cite{ByFr} contains a misprint when
$f_1$ is defined therein. The argument in the logarithm should read 
$\ln (\sqrt{\alpha^2-1}\:\mbox{dn}\, u+
\sqrt{k^2+k'^2\alpha^2}\:\mbox{sn}\, u)/
(\sqrt{\alpha^2-1}\:\mbox{dn}\, u-
\sqrt{k^2+k'^2\alpha^2}\:\mbox{sn}\, u)$.
\bibitem{note2} Calculation of the asymptotic forms of $\Pi (\phi,k)$
may be a bit tricky for the not initiated. Use of the infinite series
expansion 436.01 in \cite{ByFr} is quite helpful.
\end{thebibliography}
\end{document}